\documentclass[12pt]{article}  
\newcommand{\comma}{\;\; , \; \; }
\newcommand{\period}{\;\; .}
\newcommand{\eq}{\; = \;}
\newcommand{\sep}{\;\; , \;\;}

\newcommand{\be}{\begin{equation}}
\newcommand{\bd}{\begin{displaymath}}
\newcommand{\ee}{\end{equation}}
\newcommand{\ed}{\end{displaymath}}
\newcommand{\ba}{\begin{eqnarray}}
\newcommand{\ea}{\end{eqnarray}}
\newcommand{\trace}{{\rm trace} \; }

\newcommand{\minus}{\! - \!}
\newcommand{\mod}{{\rm mod}\,}

\renewcommand{\arraystretch}{1.5}

\title{Hard squares for  $z=-1$}

\author{ R.J. Baxter\\
{\protect \small  Mathematical
Sciences Institute,  The Australian National}\\
{\protect  \small  University,
 Canberra, A.C.T. 0200, Australia, \small e-mail: none }}

\date{\protect \small 27 September 2007}

\begin{document}

%\magnification = \magstep1
%\magnification = 1000

\maketitle
%%3456789012345678901234567890123456789012345678901234567890123456789012

 \abstract{The hard square model in statistical mechanics has been 
 investigated for the case when the activity $z$ is $-1$.  For 
 cyclic boundary conditions, the characteristic 
 polynomial of the transfer matrix has an intriguingly simple structure, 
 all the eigenvalues $x$  being zero, roots of unity, or solutions of
 $x^3 = 4 \cos^2 (\pi m/N)$.  Here we tabulate the results 
 for lattices of up to 12 columns with cyclic or free boundary 
 conditions and  the two obvious  orientations. We remark that 
 they are all unexpectedly simple and that 
 for  the rotated lattice with free or fixed boundary  conditions 
 there are obvious likely  generalizations to any lattice size.}

%%3456789012345678901234567890123456789012345678901234567890123456789012

 \vspace{5mm}

 {{\bf KEY WORDS: } Statistical mechanics, lattice models, 
 transfer matrices.}

%%%%%%%%%%%%%%%      NEW SECTION  1    %%%%%%%%%%%%%%%

%%%%%%%%%%%%%%%     Introduction       %%%%%%%%%%%%%%%

 \section{Introduction}
%%3456789012345678901234567890123456789012345678901234567890123456789012

\setcounter{equation}{0}

 The hard square lattice model has long been the subject of study. In 
 1966  Runnels and Coombs\cite{RunnelsCoombs1966} performed 
 numerical 
 calculations on lattices of up to 24 columns (an amazing  achievement 
 for  the computing power then available). Gaunt and Fisher had 
 previously  obtained high  and low-density series expansions in 
 1965\cite{GauntFisher1965}, and these were
 extended by Baxter  {\it et al} \cite{Baxteretal1980}.
 For the case $z= + 1$, Baxter also calculated the free energy to 43 
 digits of accuracy using the ``corner transfer matrix''  
 technique.\cite{Baxter1999} Recently Fernandes {\it et al} have made 
 numerical studies of various hard core two- and three-dimensional 
 systems.\cite{Fernandesetal1, Fernandesetal2}

 All these calculations naturally focussed on cases of physical 
 interest,  when the activity $z$ is real and positive, and were 
 particularly concerned  with locating and investigating  the critical 
 point at  $z \simeq 3.796$,  where the system undergoes a transition 
 from a fluid  to a solid state. 
 However, as a result of their analysis of the series expansions, Gaunt 
 and Fisher did predict singularities on the negative real axis, at  
 $z \simeq  -0.12$  and  $z \simeq  -3.80$.  Between those values the 
 free energy may have branch cuts or further singularities, and may well 
 be sensitive to the boundary conditions imposed.

%%3456789012345678901234567890123456789012345678901234567890123456789012

 Fendley {\it et al} \cite{Fendleyetal2005}  studied the case when 
 $z=-1$, using the usual cyclic boundary conditions and orientation. 
 For lattices of $N$ columns, with $N \leq 15$,  they found that the 
 characteristic polynomial $P_N(x) = \det (x I - T)$ of the 
 row-to-row transfer matrix is 
 remarkable simple, being a product of factors of the form 
 $x^m \pm 1$. This implies that all the eigenvalues of $T$ lie on 
 the unit circle. They conjectured that this behaviour
 held for lattices of any width.
 
 This conjecture was verified by Jonsson,\cite{Jonsson2006} who 
 established an equivalence between the model and the counting of 
 particular rhombus tilings of the plane. Moreover, he obtained an 
 algorithm for counting these tilings that grows only polynomially 
 (instead of exponentially) with  the size of the lattice.
 He was therefore able to calculate $P_N(x)$ 
 explicitly for lattices of up to 50 columns.
 
 More recently, Jonsson \cite{Jonsson2006a}  has 
 investigated the effect of first rotating
 the lattice through $45^{\circ}$ before imposing toroidal
 boundary conditions, and Bousquet-M{\'e}lou {\it et al}\cite{Bousquet2007}
 have looked at the problem using free boundary conditions.
 Again they found that $P_N(x)$ is a product of 
 simple factors. 
 
  We have explicitly calculated $P_N(x)$ for $N \leq 12$, and for 
  completeness present the results here for the three cases 
  mentioned, plus  the remaining case of the $45^{\circ}$ orientation 
  with free (or fixed) boundary conditions. This last case is 
  particularly simple, leading us to conjecture its  form for
  general $N$.
 %%3456789012345678901234567890123456789012345678901234567890123456789012

 \section{The $90^{\circ}$ orientation}
 \setcounter{equation}{0}

 Fendley {\it et al} \cite{Fendleyetal2005} considered the square 
 lattice 
 $\cal L$ drawn in the obvious way as in Figure \ref{sqlatt}, with $M$ 
 rows and $N$ columns and $R = M N$ sites.  For the hard squares model, 
 with each site $i$ is associated an occupation number $\sigma_i$, 
 which takes the values $0$ or $1$, corresponding to whether the site 
 is empty or contains a particle. At most one particle can lie on any 
 site.

%%3456789012345678901234567890123456789012345678901234567890123456789012

%% The following picture is allowed a space 14cm by 6cm
 \setlength{\unitlength}{1pt}
 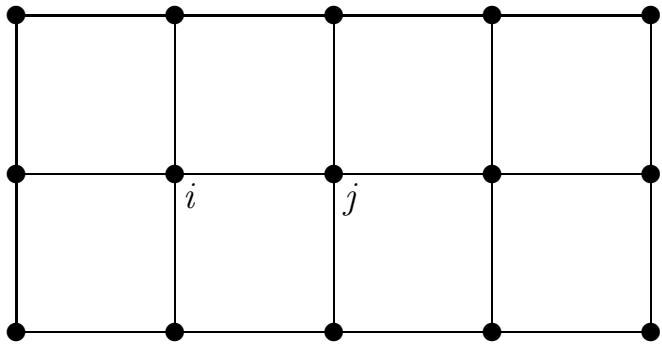
\begin{figure}[hbt]
 \begin{picture}(420,160) (0,0)

 \multiput(60,0)(60,0){5}{\circle*{7}}
 \multiput(60,60)(60,0){5}{\circle*{7}}
 \multiput(60,120)(60,0){5}{\circle*{7}}

 \put (60,0) {\line(1,0) {240}}
 \put (60,60) {\line(1,0) {240}}
 \put (60,120) {\line(1,0) {240}}
 \put (60,0) {\line(0,1) {120}}
 \put (120,0) {\line(0,1) {120}}
 \put (180,0) {\line(0,1) {120}}
 \put (240,0) {\line(0,1) {120}}
 \put (300,0) {\line(0,1) {120}}

 \put (123,47) {\large \it i}
 \put (184,47) {\large \it j}

 \end{picture}
 \vspace{1.5cm}
 \caption{\footnotesize The square lattice of solid lines and 
 circles,  showing two adjacent sites $i$ , $j$.}
 \label{sqlatt}
 \end{figure}
 
%%3456789012345678901234567890123456789012345678901234567890123456789012

 Particles are also not allowed to occupy adjacent sites. We can 
 express this by defining a weight function $W(\sigma_i,\sigma_j)$ for 
 each edge $\langle i,j \rangle$ of $\cal L$  by
 \ba W(a,b) &\!  =   \! & 1  \; \; {\rm if} \;\;  a, b \; \; 
 {\rm not \; both} \; 1 \comma \\
 &  \! =  \! &  0\; \; {\rm if} \;\;  a = b= 1  \period \ea
 
 Let { \boldmath${\sigma}$}$= \{ \sigma_1, \ldots, \sigma_R \}$  denote the 
 set of the occupation numbers of all the 
 sites of $\cal L$, and let
 \be \label{spinsum}
 n({\mbox{\boldmath$\sigma$}}) = \sum_i \sigma_i  \comma \ee
 the sum being over all the $R$ sites $i$. Then the partition function is
 \be \label{ptfn}
 Z_{M,N} \eq \sum_{{\mbox{\boldmath$\sigma$}}} 
 z^{{\textstyle n}({\mbox{\boldmath$\sigma$}})} \; \prod_{\langle i,j \rangle} 
 W(\sigma_i, \sigma_j) \comma \ee
 the product being over all edges ${\langle i,j \rangle}$ of $\cal L$ 
 and the sum over all values of 
 $\sigma_1, \sigma_2, \ldots , \sigma_R$.
 
 Define the partition function per site
 \be \label{defkappa}
 \kappa = \kappa (z) = (Z_{M,N})^{1/R} \period \ee
 The parameter $z$ is known as the {\it activity}. For real 
 positive values of $z$ we expect $\kappa(z)$ to tend to a 
 positive limit as  $M, N  \rightarrow \infty$, to be independent of 
 the order in which this occurs, and  indeed of the boundary 
 conditions imposed.   The free energy or entropy is 
 proportional to $- \log \kappa(z)$ and  is the function investigated by 
 Runnels and  Coombs,\cite{RunnelsCoombs1966} Gaunt and 
 Fisher,\cite{GauntFisher1965}  and Baxter 
 {\it et al} \cite{Baxteretal1980}.

%%3456789012345678901234567890123456789012345678901234567890123456789012

 If we impose toroidal boundary conditions, so that row $M$ is followed 
 by row 1,  and column $N$ by column 1, then it is straightforward to 
 show that
 \be \label{Ztr}
 Z_{M,N} = \trace (T_N)^M \comma \ee
 where $T_N$ is the {\em transfer matrix}, with entries
 \be \label{transmat}
 (T_N)_{\sigma, \sigma'}\eq \prod_{j=1}^N z^{\sigma_j} 
 \, W(\sigma_j, \sigma_{j+1}) W(\sigma_j, \sigma'_j) \ee
 where now $\sigma = \{ \sigma_1, \ldots ,\sigma_N \}$,
 $\sigma' = \{ \sigma'_1, \ldots ,\sigma'_N \}$  and 
 $\sigma_{N+1} = \sigma_1$.
 The RHS is the combined weight of two adjacent rows of $\cal L$, 
 with occupation numbers $\sigma_1, \ldots ,\sigma_N$ in one row, 
 and $\sigma'_1, \ldots ,\sigma'_N$ in the row above, as in 
 Figure \ref{tworowss}.

%%3456789012345678901234567890123456789012345678901234567890123456789012

 %% The following picture is allowed a space 14cm by 6cm
 \setlength{\unitlength}{1pt}
 \begin{figure}[hbt]
 \begin{picture}(420,160) (0,0)

 \multiput(60,0)(60,0){5}{\circle*{7}}
 \multiput(60,60)(60,0){5}{\circle*{7}}

 \put (60,0) {\line(1,0) {240}}
 \put (60,60) {\line(1,0) {240}}
 \put (60,0) {\line(0,1) {60}}
 \put (120,0) {\line(0,1) {60}}
 \put (180,0) {\line(0,1) {60}}
 \put (240,0) {\line(0,1) {60}}
 \put (300,0) {\line(0,1) {60}}

 \put (47,-13) {$\sigma_1$}
 \put (107,-13) {$\sigma_2$}
 \put (287,-13) {$\sigma_N$}
 \put (45,47) {$\sigma'_1$}
 \put (105,47) {$\sigma'_2$}
 \put (283,47) {$\sigma'_N$}

 \end{picture}
 \vspace{1.5cm}
 \caption{\footnotesize Two rows of $\cal L$, corresponding to the 
 element $ T_{\sigma, \sigma'}$  of the transfer   matrix $T$.}
 \label{tworowss}
 \end{figure}
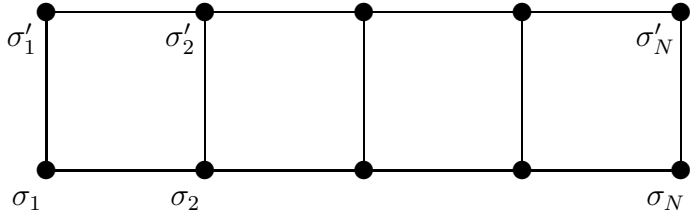
 
 %%3456789012345678901234567890123456789012345678901234567890123456789012
 
 As defined here, $T_N$ is a $2^N$ by $2^N$ dimensional matrix.
 However, many of its rows vanish, corresponding to two horizontally
 adjacent occupation numbers $\sigma_j, \sigma_{j+1}$ being both unity.
 It is natural to exclude such rows and to impose the condition
 \be \label{restr}
 \sigma_j, \sigma_{j+1} \; \; {\rm not \; both \; one,} \; \;  
 j = 1,\ldots , N  \ee
 on both the sets $\sigma$ and $\sigma'$. This does not change the RHS of 
 (\ref{Ztr}). If $S$ is the two-by-two matrix
 \renewcommand{\arraystretch}{1.0}
  
 \bd  S = \left(  {\scriptscriptstyle  
{ \renewcommand{\arraycolsep}{3pt}
\begin{array}{ cc}
     1 & 1   \\
    1  &  0 
 \end{array} } }\right)   \comma \ed
 then the  matrix $T_N$ is  of dimension
 \bd  D_N =  \trace S^N \period \ed

   \renewcommand{\arraystretch}{1}

 The characteristic polynomial of the matrix $T_N$ is 
 \be\label{defPN}
 P_N(x) = \det (x \, I -T_N) \period \ee
 For arbitrary values of the activity $z$,  one would not 
 expect this to have any simple structure. However, 
 the case
 \be z= -1 \ee  seems to be very special. Fendley 
 {\it et al} \cite{Fendleyetal2005} evaluated
 $P_N(x)$ for $N= 1, \ldots ,15$ and found that 
 it to be amazingly simple, being in each case a product of 
 factors of the form
 \bd (x^n \pm 1) \period \ed
 Thus all the eigenvalues of $T_N$ lie on the unit circle.

 \begin{table}[hbt]
 \begin{center}
 \begin{tabular} {| c | c | c |}
 \multicolumn{3}{c} {} \\
 \hline
 $ N $  & $ D_N$  & $ P_N (x)$ \\ \hline
  {\small  1  } &   {\small   1 } & $f_1  $ \\
   {\small  2 } &  {\small   3 } & $f_1 f_4/f_2 $ \\
   {\small   3 } &   {\small  4 } & $f_1 f_3$ \\
   {\small  4 } &  {\small 7 } & $ f_1 f_2 f_4$ \\
    {\small  5 } &   {\small  11 } &  $f_1 
    f_{10}^{\,2} /f_5^{\,2}$  \\
    {\small  6 } &   {\small 18 } & $f_1
     f_3 f_4 f_{6}^{\,2}/f_2 $ \\
    {\small  7 } &  {\small  29 } & $ f_1 
    f_{28}^{\, 2}/f_{14}^{\, 2} $ \\
    {\small  8 } &  {\small 47 } & 
    $f_1 f_2 f_4 f_{10}^{\,4}$ \\ 
    {\small    9 } &  {\small 76 } &  $f_1 
    f_3 f_{9}^{\, 4} f_{18}^{\, 2} $ \\
    {\small   10 } &  {\small 123 } &  $f_1
     f_4 f_5^{\, 2} f_{8}^{\, 5} f_{14}^{\, 5}
      /f_2 $ \\ 
   {\small  11 } &  {\small 199  } &  $f_1 
    f_{44}^{\, 4} f_{55}^{\, 2}/f_{22}^{\, 4}
       $ \\
   {\small  12 } &  {\small 322 }  &  $f_1
    f_2 f_3  f_4 f_{6}^{\, 2} f_{12}^{\, 12} 
   f_{18}^{\, 6} f_{24}^{\, 2}  $ \\
 \hline
 \end{tabular}
 \end{center}
 \caption{\small $P_N(x)$ for the $90^{\circ}$ orientation
  with cyclic boundary conditions.}
 \vspace{0.5cm}
 \label{90degcyclic}
\end{table}

%%3456789012345678901234567890123456789012345678901234567890123456789012

 We define the functions
 \be f_n \eq f_n(x) \eq x^n-1  \comma \ee
 and give Fendley {\it et al}'s  results 
 up to $N=12$ in
 Table \ref{90degcyclic}.  In each case the denominator is a factor of 
 the numerator, so each expression is 
 in fact a polynomial in $x$, of  degree $D_N$.

%%3456789012345678901234567890123456789012345678901234567890123456789012

 \subsection*{Free and fixed boundary conditions}
 One does not have to impose toroidal (i.e. cyclic in both the vertical 
 and horizontal  directions)  boundary conditions on the 
 lattice $\cal L$. Other boundary conditions are of interest in 
 statistical mechanics, in particular cylindrical boundary conditions, 
 where $\cal L$ begins on the left at column 1 and ends on the 
 right at column N. If one still imposes cyclic boundary conditions 
 from top to bottom (so that row $M$ is followed by row 1), then 
 $Z_{M,N}$ is again given by (\ref{Ztr}) and  (\ref{transmat}), except 
 that the single factor $W(\sigma_N,\sigma_{N+1})$ is omitted.
 
%%3456789012345678901234567890123456789012345678901234567890123456789012
 
 This is equivalent to requiring that (\ref{restr}) hold only for
 $j = 1, \ldots, N\minus 1$, so now $T_N$ is a square matrix of 
 dimension
 \bd D_N = \sum_{i=1}^2  \sum_{j=1}^2 (S^{N-1})_{i,j} \period  \ed
 
 This case is discussed by Bousquet-M{\'e}lou 
 {\it et al}\cite{Bousquet2007}, who calculate $P_N(x)$ for 
 $N=1,\ldots , 10$.
 We have extended these calculations to $N=1,\ldots , 12$ 
 and give
 the results in Table \ref{90degfree}.
 Again the denominators are factors of the numerators, so each 
 expression is  a polynomial, for instance
 \bd P_1(x) = x^2-x+1 \sep P_2(x) =  (x-1) (x^2+1) \period \ed
 
 \begin{table}[hbt]
 \begin{center}
 \begin{tabular} {| c | c | c |}
 \multicolumn{3}{c} {} \\
 \hline

 $ N $  & $ D_N$  & $ P_N (x)$ \\ \hline
    \small 1  &   \small 2 & $f_1 f_6/(f_2 f_3) $ \\
   \small  2 &  \small 3 & $f_1 f_4/f_2 $ \\
    \small 3 &  \small 5 & $f_1 f_8/f_4$ \\
    \small 4 &  \small 8 & $ f_1 f_4 f_6/f_3$ \\
    \small 5 &   \small 13 &  $f_1 f_8 f_{10} /(f_2 f_4)$  \\
    \small 6 &   \small 21 & $f_1 f_4^{\, 2}  f_{14}/f_2 $ \\
   \small  7 &   \small 34 & $ f_1 f_6 f_8 f_{12}
    f_{18}/(f_3 f_4^{\, 2}) $ \\
    \small 8 &  \small 55 & $f_1 f_4^{\, 2}
     f_{16}^{\,2}f_{22}/f_8$ \\ 
    \small 9 &  \small 89 &  $f_1 f_8 f_{14} 
    f_{20}^{\, 3} f_{26}/(f_2 f_4^{\, 2} 
     f_{10} )$ \\
  \small  10 & \small144 &  $f_1 f_4^{\, 3} f_6 
  f_{18}^{\, 2} f_{24}^{\, 3}
      f_{30}/(f_2 f_3 f_8)$ \\ 
   \small 11 & \small233  &  $f_1 f_8 f_{14} f_{16}
    f_{22}^{\, 4} f_{28}^{\, 3}
       f_{34}/f_4^{\, 3} $ \\
   \small 12 & \small377 &  $f_1 f_4^{\, 3}f_{10} 
   f_{20}^{\, 2} f_{26}^{\, 6} 
         f_{32}^{\, 4} f_{38}/f_8 $ \\
 \hline
 \end{tabular}
 \end{center}
 \caption{\small $P_N(x)$ for the $90^{\circ}$ orientation
  with free boundary conditions.}
 \vspace{0.5cm}
 \label{90degfree}
 \end{table}
%%3456789012345678901234567890123456789012345678901234567890123456789012

 We observe that, surprisingly, these results are similar in form to 
 those for the cyclic case, in particular all the eigenvalues, i.e. 
 the zeros of $P_N(x)$, lie on the unit circle.

 We can also consider {\em fixed boundary conditions}, for instance
 we could fix $\sigma_1, \sigma_N, \sigma'_1, \sigma'_N$ in  
 (\ref{transmat}) to be zero. (The factor $W(\sigma_N, \sigma_1)$
 is then unity, so it is irrelevant whether it is removed.)
 However, since then $W(\sigma_1, \sigma_2) = W(\sigma_{N-1}, \sigma_N) 
 =1 $ for all $\sigma_2$ and $\sigma_{N-1}$, this is merely the same as 
 imposing free boundary conditions on  columns 2 and $N\minus 1$ of 
 $\cal L$, so for the $90^{\circ}$ orientation
 \be
 \left[ P_N(x) \right]_{\rm fixed} = \left[ P_{N-2}(x) \right]_{\rm free} 
 \period \ee

 \section{The $45^{\circ}$ orientation}
 \setcounter{equation}{0}
 We now consider the square lattice $\cal L$ turned through 
 $45^{\circ}$, as in  Figure \ref{sqlatt45}.

 %% The following picture is allowed a space 14cm by 6cm
 \setlength{\unitlength}{1pt}
 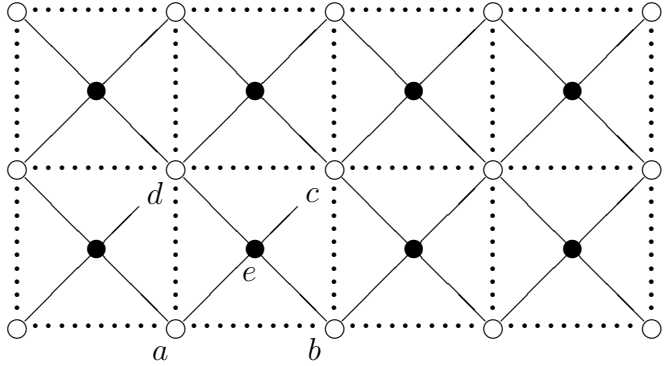
\begin{figure}[hbt]
 \begin{picture}(420,160) (0,0)

 \put (63,63) {\line(1,1) {54}}
 
 \put (123,63) {\line(1,1) {54}}
 \put (183,63) {\line(1,1) {54}}
 \put (243,63) {\line(1,1) {54}}
 \put (63,3) {\line(1,1) {43}}
 \put (123,3) {\line(1,1) {43}}
 \put (183,3) {\line(1,1) {54}}
 \put (243,3) {\line(1,1) {54}}
 
 \put (63,117) {\line(1,-1) {54}}
 \put (123,117) {\line(1,-1) {54}}
 \put (183,117) {\line(1,-1) {54}}
 \put (243,117) {\line(1,-1) {54}}
 
 \put (63,57) {\line(1,-1) {54}}
 \put (123,57) {\line(1,-1) {54}}
 \put (183,57) {\line(1,-1) {54}}
 \put (243,57) {\line(1,-1) {54}}

 \multiput(60,0)(60,0){5}{\circle{7}}
 \multiput(60,60)(60,0){5}{\circle{7}}
 \multiput(60,120)(60,0){5}{\circle{7}}
 \multiput(90,30)(60,0){4}{\circle*{7}}
 \multiput(90,90)(60,0){4}{\circle*{7}}

 \multiput(65,0)(5,0){10}{\bf .}
 \multiput(125,0)(5,0){10}{\bf .}
 \multiput(185,0)(5,0){10}{\bf .}
 \multiput(245,0)(5,0){10}{\bf .}
 \multiput(65,60)(5,0){10}{\bf .}
 \multiput(125,60)(5,0){10}{\bf .}
 \multiput(185,60)(5,0){10}{\bf .}
 \multiput(245,60)(5,0){10}{\bf .}
 \multiput(65,120)(5,0){10}{\bf .}
 \multiput(125,120)(5,0){10}{\bf .}
 \multiput(185,120)(5,0){10}{\bf .}
 \multiput(245,120)(5,0){10}{\bf .}

 \multiput(58,7)(0,5){10}{\bf .}
 \multiput(58,67)(0,5){10}{\bf .}
 \multiput(118,7)(0,5){10}{\bf .}
 \multiput(118,67)(0,5){10}{\bf .}
 \multiput(178,7)(0,5){10}{\bf .}
 \multiput(178,67)(0,5){10}{\bf .}
 \multiput(238,7)(0,5){10}{\bf .}
 \multiput(238,67)(0,5){10}{\bf .}
 \multiput(298,7)(0,5){10}{\bf .}
 \multiput(298,67)(0,5){10}{\bf .}
 
 \put (111,-12) {$a$}
 \put (170,-12) {$b$}
 \put (109,47) {$d$}
 \put (169,48) {$c$}
 \put (145,18) {$e$}
 
 \end{picture}
 \vspace{1.5cm}
 \caption{\footnotesize The square lattice $\cal L$  turned 
 through $45^{\circ}$, indicated by all the circles and the 
 solid lines, and the reduced lattice ${\cal L}_1$ shown by 
 open circles and dotted lines. }
 \label{sqlatt45}
 \end{figure}

%%3456789012345678901234567890123456789012345678901234567890123456789012

 If we sum over the occupation numbers on alternate rows (the 
 solid circles  
 in the figure), then we obtain a  new square lattice ${\cal L}_1$, 
 shown by the open circles and dotted lines in 
 Figure \ref{sqlatt45}, which has the 
 usual $90^{\circ}$ orientation used in the previous section. The four 
 sites round a face  of ${\cal L}_1$, e.g. those with occupation 
 numbers $a, b, c, d $ shown, have the weight
 \be {W}_1 (a, b, c, d) \eq \sum_{e=0}^1 
 z^e \,  W(a,e)  W(b,e) W(c,e) W(d,e) \period \ee
 Thus
 \ba  {W}_1 (a, b, c, d) & \! = \! & 1+z \; \; 
 {\rm if \; } a=b=c=d=0 \comma \\
 & \! = \! & 1 \;\;\;\;\; {\rm otherwise} \ea
 and we see that $z=-1$ is special in the sense 
 that $W_1(0,0,0,0)$ then vanishes.

 The definition  (\ref{ptfn}) is equivalent to 
 \be
 Z_{M,N} \eq \sum_{{\mbox{\boldmath$\sigma$}}} 
 z^{{\textstyle n}({\mbox{\boldmath$\sigma$}})} \, 
 \prod {W}_1(\sigma_i,\sigma_j,
 \sigma_k,\sigma_l) 
 \comma \ee
 the product being over all faces $(i,j,k,l)$ of  ${\cal L}_1$ and 
 the outer sum over all values of the occupation  numbers 
 ${\mbox{\boldmath$\sigma$}}$ on 
 ${\cal L}_1$. Again $n({\mbox{\boldmath$\sigma$}})$ is 
 defined by (\ref{spinsum}), 
 but now the sum is over all sites of  ${\cal L}_1$. We now take
 $M, N, R$ to be the number of rows, columns and sites of
 ${\cal L}_1$, so again  $R = M N$.

%%3456789012345678901234567890123456789012345678901234567890123456789012

 The transfer matrix is no longer given by (\ref{transmat}), but by
 \be \label{transmat2} (T_N)_{\sigma, \sigma'}\eq
 \prod_{j=1}^N z^{\sigma_j} \, 
 W_1(\sigma_j, \sigma_{j+1},\sigma'_{j+1}, \sigma'_j) \ee
 and is of dimension $2^N$. None of the rows or columns now vanish,
 but many of the rows of $T_N$ are  equal or opposite. This is because
 two rows with $\sigma = \{a, \ldots ,b, 1,0,1,c,\ldots , d\}$ and 
 $\{a, \ldots ,b, 1,1,1,c,\ldots, d\}$ must be equal and opposite.
 (The central occupation number is different, but the face weights
 on either side are 1 for each of the two cases.)
 Thus many of the eigenvalues of $T_N$ will be zero, and we do  
 observe this.

 For positive integers $n$, define
 \be g_n = g_n(x) = x^3-n \period \ee
 If we impose cyclic boundary conditions,  so that column $N$ of 
 ${\cal L}_1$ is followed by column 1, then for $z=-1$ and 
 $N= 1, \ldots , 12$ we find that the characteristic polynomial 
 $P_N(x)$, as defined by (\ref{defPN}), is given as in Table
 \ref{45degcyclic}.

%%3456789012345678901234567890123456789012345678901234567890123456789012

 \begin{table}[hbt]
 \begin{center}
 \begin{tabular} {| c | c | c |}
 \multicolumn{3}{c} {} \\
 \hline

 $ N $  & $ D_N$  & $ P_N (x)$ \\ \hline
   \small 1  &  \small 2 & $f_3/f_1 $ \\
   \small 2 & \small 4 & $x^2 f_3/f_1 $ \\
   \small 3 & \small 8 & $x^{\, 3} f_1^{\, 2} 
   g_4 $ \\
   \small 4 & \small 16 & $x^6 f_3
    f_4^{\, 2}/f_1$ \\
   \small 5 &  \small 32 &  $x^{20} f_3
    f_5^{\, 2}/f_1$  \\
   \small 6 &  \small 64 & $x^{44} f_1^{\,2} 
   g_1^{\,2} g_3^{\,2} g_4^{\,2}$ \\
   \small 7 &  \small 128 & $ x^{84} f_3 
   f_{21}^{\, 2}/f_1 $ \\
   \small 8 & \small 256 & $ x^{198} f_3 
   f_4^{\, 2} f_{24}^{\, 2}/f_1 $ \\ 
  \small  9 & \small 512 &  $x^{408} f_1^{\,2} 
  g_1^{\, 12} g_4^{\, 4} \, 
       (x^9-6x^6+9 x^3-1)^6 $ \\
  \small 10 & \small 1024 &  $x^{832} f_3 
  f_5^{\, 2} f_{30}^{\, 6}/f_1$ \\ 
  \small 11 & \small 2048  &  $x^{1782} 
  f_3 f_{33}^{\, 8}/f_1 $ \\
  \small 12 & \small 4096 &  $x^{3660} 
  f_1^{\, 2} f_4^{\, 2} g_1^{\, 66}  
    g_2^{\, 16}  g_3^{\, 18}  g_4^{\, 10} \,
       (x^6-4 x^3+1)^{16} $ \\
 \hline
 \end{tabular}
 \end{center}
 \caption{\small $P_N(x)$ for the $45^{\circ}$ orientation
  with cyclic boundary conditions.}
 \vspace{0.5cm}
 \label{45degcyclic}
 \end{table}

%%3456789012345678901234567890123456789012345678901234567890123456789012

 If $N$ is divisible by 3, we see that there are extra factors
 whose zeros do not in general lie on the unit circle.
 Nevertheless they do have a simple structure: in each case 
 $P_N(x)$ is a product (or ratio) of factors $x, f_n$ and of
 \be \label{smn}
 s(m,N) = x^3 - 4 \cos^2 (\pi m/N)    \period \ee
 To see this, note that
 $g_4 = s(0,N)$, $g_3 = s(N/6,N)$, $g_2 = s(N/4,N)$,  
 $g_1 = s(N/3,N)$ , and
 \bd 
 x^9-6x^6+9 x^3-1 = s(1,9) \, s(2,9) \, s(4,9) \comma \ed
 \bd 
 x^6-4 x^3+1 = s(1,12) \, s(5,12)  \period \ed
 This is the case investigated by Jonsson in his second 
 paper\cite{Jonsson2006a}. Using the equivalence to 
 rhombus tilings of the plane, he does indeed show that 
 such factors $s(m,N)$ occur when $N$ is divisible by three.
 
 Thus the eigenvalues of $T_N$ lie either at the origin,
 on the unit circle, or their cubes lie on the interval (0,4]
 of the positive real axis.

 \subsection*{Free and fixed boundary conditions}
 As with the $90^{\circ}$ case, we can instead impose free
 boundary conditions, so the factor
 $W_1(\sigma_N,\sigma_1,\sigma'_1,\sigma'_N)$ (occurring 
  when $j=N$)
 in the product in  (\ref{transmat2}) is omitted. Now we  find, 
 for $N= 1, \ldots, 12$ that $P_N(x)$ has the very simple forms 
 given in Table  \ref{45degfree}.

 \begin{table}[hbt]
 \begin{center}
 \begin{tabular} {| c | c | c |}
 \multicolumn{3}{c} {} \\
 \hline

 $ N $  & $ D_N$  & $ P_N (x)$ \\ \hline
   {\small 1}  &  {\small 2} & $x^2 $ \\
   {\small 2} & {\small 4} & $x^2 f_3/f_1 $ \\
   {\small 3} & {\small 8} & $x^4 f_1 f_3 $ \\
  { \small 4} & {\small 16} & $x^{16} $ \\
   {\small 5} &  {\small 32} &  $x^{24} f_3^{\, 3} /f_1$  \\
  { \small 6} &  {\small 64} & $x^{48} f_1 f_3^{\, 5} $ \\
   \small 7 &  \small 128 & $ x^{128}  $ \\
   \small 8 &\small  256 & $ x^{224} f_3^{\, 11}/f_1 $ \\ 
   \small 9 & \small 512 &  $x^{448} f_1 f_3^{\,21}  $ \\
  \small 10 & \small 1024 &  $x^{1024} $ \\ 
  \small 11 & \small 2048  &  $x^{1920} f_3^{\,43}/f_1 $ \\
  \small 12 & \small 4096 &  $x^{3840} f_1 f_3^{\, 85} $ \\
 \hline
 \end{tabular}
 \end{center}
 \caption{\small $P_N(x)$ for the $45^{\circ}$ orientation
  with free boundary conditions.}
 \vspace{0.5cm}
 \label{45degfree}
 \end{table}

%%3456789012345678901234567890123456789012345678901234567890123456789012

 For a given $N$, define $m $ by
 \ba m= &  (4^k-1)/3 & {\rm if \; } N=3k  \\
 = &   0  &{\rm if \; }  N= 3k+1 \\
 = & \!   (1+ 2^{2k+1})/3  &{\rm if \; } 
 N=3k+2 \comma \ea
 where $k$ is an integer. Also set
 \bd  r = 1- \mod (N,3)  \sep n = 2^N-3m -r  \ed
 (so $r = 1, 0$ or $-1$).
 Then the results of Table \ref{45degfree} are fitted by
 \be \label{45free}
 P_N(x) = \left[ P_{N}(x) \right]_{\rm free} = x^n f_3^m f_1^r 
 \period \ee

%%3456789012345678901234567890123456789012345678901234567890123456789012

 As with the $90^{\circ}$ orientation, we can also consider 
 {\em fixed boundary conditions}, in particular by  fixing the 
 occupation numbers in the first and last columns of ${\cal L}_1$
 to be zero. However, if we go back to the original lattice $\cal L$
 (the solid lines in Figure \ref{sqlatt45})  which has $2N-1$ columns, 
 this is equivalent to imposing free boundary conditions on 
 columns 2 and $2N-2$. This means that 
 \be
 \left[ Z_{M,N} \right]_{\rm fixed} = \left[ Z_{M,N-1} \right]_{\rm free}
 \period \ee
 The same relation does not quite hold for the transfer matrices,
 because $T_N$ for fixed boundary conditions is of dimensions
 $d$, while  $T_{N-1}$ for free boundaries is of dimensions
 $2 d$, where
 \be d= 2^{N-2} \period \ee
 For fixed or free boundary conditions we see from
 Figure \ref{sqlatt45} that there are two sorts of rows of the 
 original lattice $\cal L$: the ones shown as open circles containing 
 $N$ sites per row, and the ones shown as solid circles 
 containing only $N\minus 1$ sites. This ensures 
 that the transfer matrices factor:
 \be 
  \left[ T_{N \minus 1} \right]_{\rm free} = S_N R_N  \sep  
  \left[ T_N \right]_{\rm fixed} = R_N S_N \comma \ee
 where  $R_N$ is a $d$ by $2 d$ matrix  and
 $S_N$ is $2 d$ by $d$. This means that
 $d$ of  the eigenvalues of  $T_{N-1}$(free) vanish trivially, while 
 the other $d$ are the same as those of   $T_{N}$(fixed). Hence
 \be \label{45fixed}
 \left[ P_N(x) \right]_{\rm fixed} = 
 \left[ P_{N-1}(x) \right]_{\rm free}/x^{d }
 \period \ee

 Thus for both orientations the characteristic polynomial
 for fixed boundary conditions is obtainable from that for free
 boundary conditions.

%%3456789012345678901234567890123456789012345678901234567890123456789012

 \section{Conclusions}
\setcounter{equation}{0}

For cyclic boundary conditions, with either the $45^{\circ}$ or 
$90^{\circ}$ orientations, Jonsson has shown that $P_N(x)$ 
is product or ratio of factors of the form $x, x^m-1$, and (for the
 $45^{\circ}$ case with $N$ divisible by three) 
 $x^3-4 \cos^2(\pi m/N)$. Somewhat remarkably, these 
 properties appear to remain true when we impose free or 
 fixed boundary conditions. In particular, we conjecture that
  for the $45^{\circ}$ free or fixed cases, $P_N(x)$ is a product or 
 ratio only of the simple factors $x, f_1(x), f_3(x)$, being  
 given by (\ref{45free}) and (\ref{45fixed}).

%%3456789012345678901234567890123456789012345678901234567890123456789012
 For the $90^{\circ}$  cyclic case, if we take the limit when $M,N$ become 
 large while remaining co-prime, then $Z_{M,N} = 1$. On the other
 hand, from our conjecture (3), for the $45^{\circ}$
 free case, if $N= 3k+1$, then $r,m$ in  (\ref{45free}) are both zero,
 giving
 \be Z_{M,N} = 0  \period \ee

 {From } (\ref{defkappa})  $\kappa$ for these two cases is 1 and 0, so 
 its limit is certainly dependent on the boundary conditions. This can 
 happen, because when $z=-1$ the Boltzmann weights are not all 
 positive.

 So $z=-1$ is a very special non-physical case of the hard squares
 model. Even so, the transfer matrix does then appear to have very  
 intriguing properties.
 
  \section{Acknowledgement}
  
  The author thanks Paul Fendley for very helpful correspondence, in 
  particular for bringing the Web reports of the recent work by 
   Jonsson and Bousquet-M{\'e}lou {\it et al} to his attention.
  
\setcounter{equation}{0}

%%3456789012345678901234567890123456789012345678901234567890123456789012

 \end{document}